\documentclass[aps, prl,amsmath,amssymb,twocolumn,floatfix]{revtex4}
\usepackage{graphicx}
\usepackage{bm}
\usepackage{comment}
\usepackage{epsfig}

\newcommand{\be}{\begin{equation}} 
\newcommand{\ee}{\end{equation}}
\newcommand{\bea}{\begin{eqnarray}} 
\newcommand{\eea}{\end{eqnarray}}

\begin{document}
\title{\bf A Monte Carlo method for chemical potential determination in single and multiple occupancy crystals}
%\title{\bf Lattice size switching: Monte Carlo determination of chemical potentials in crystals and cluster solids}
%\title{\bf Direct determination of crystal chemical potentials and equilibrium conditions in cluster crystals}
\author{Nigel B. Wilding}
\affiliation{Department of Physics, University of Bath, Bath BA2 7AY, United Kingdom.}

\author{Peter Sollich}
\affiliation{King's College London, Department of Mathematics, Strand,
London WC2R 2LS, United Kingdom.}

%\date{\today}

\begin{abstract}

We describe a Monte Carlo scheme which, in a single simulation, yields a
measurement of the chemical potential of a crystalline solid. Within the
isobaric ensemble, this immediately provides an estimate of the system
free energy, with statistical uncertainties that are determined
precisely and transparently. An extension to multiple occupancy
(``cluster'') solids permits the direct determination of the cluster
chemical potential and hence the equilibrium conditions. We apply the
method to a model exhibiting cluster crystalline phases, where we find
evidence for an infinite cascade of critical points terminating
coexistence between crystals of differing site occupancies.

\end{abstract} 
\maketitle

The phase behaviour of crystalline materials is important in fields as
diverse as solid state physics, soft matter, mineralogy and
pharmacology. For instance, metals and their alloys exhibit rich phase
behaviour \cite{Young:1991} (including novel features such as
isostructural transitions \cite{Decremps:2011vn}); colloids can
self assemble into a variety of complex structures with applications in
photonics \cite{Ye:2001uq};  many drug compounds exhibit crystalline
polymorphism which can influence their clinical function
\cite{drugbook}. The staple simulation approach for predicting
crystalline phase behaviour is via free energy estimates obtained by
numerical integration along some path that connects the macrostate of
interest to a reference state of known free energy
\cite{Frenkel1984,Frenkelsmit2002}. Such ``Thermodynamic Integration''
(TI) is popular because it is both conceptually simple and can be
implemented with only a modest extension of the simulation framework
needed for standard Monte Carlo (MC) sampling. However there are a
number of respects in which it is less than ideal. The method hinges on
the identification of a good path and reference macrostate. A `good'
path is short; but the reference macrostate (the choice of which is
limited) may lie far from the physical macrostate of interest, entailing
a large number of independent simulations to make the necessary link. A
potentially more serious constraint on the path is that the derivative
being measured should vary slowly, smoothly and reversibly along it; if
it does not the numerical quadrature may be compromised. A phase
transition en route is thus a particular hazard. Evidently one has to
decide how many simulations are to be performed along the path and
where. In so doing one must strike a suitable balance between minimizing
computation time and ensuring that no region of the path  is neglected.
This may necessitate a degree of trial and error. The uncertainties to
be attached to TI estimates are also problematic. Use of simple
numerical quadrature will result in errors. Error bounds have to
aggregate the uncertainties (statistical and systematic) associated with
different stages of the integration process.

Beyond simple crystals, there is considerable interest in particles that
self assemble via microphase separation into periodically modulated
nanostructures. Classic examples include the lamellar and micellar
crystalline phases encountered in surfactants and copolymers
\cite{Alexandridis:2000}. More recently, it has been discovered that
when certain types of repulsive particles that lack a hard repulsive
core are compressed to high density, multiple occupancy (``cluster'')
crystals are formed 
\cite{Mladek:2006ys,Mladek2007,Zhang:2010fk,Neuhaus:2011fk}. Such
coreless potentials serve as models for a wide range of soft matter
systems such as star polymers, dendrimer and microgels in which
particles can substantially overlap \cite{Likos:2006}.  To describe
cluster crystals one must allow for a crystalline lattice in which each
lattice site can be occupied by multiple particles. Let us suppose that
such a crystal has $N_{\rm c}$ lattice sites, labeled $i=1...N_{\rm c}$ and that
site $i$ is occupied by $n_{\rm c}(i)$ particles (a ``cluster''). Clusters are
generally bi- or poly-disperse, so the total particle number is
$N=\sum_{i=1}^{N_{\rm c}}n_{\rm c}(i)=N_{\rm c}n_{\rm c}$, with $n_{\rm c}$ the average occupancy. A
particular problem for simulation is to determine the equilibrium values
of $n_{\rm c}$, the lattice parameter $a$, the pressure $P$ and the chemical
potential $\mu$ that correspond to some particle number density
$\rho=N/V$ and temperature $T$ of interest. As shown previously,
measurement of the Helmholtz free energy $F$ at fixed $N_{\rm c}$ in the
constant-($NVT$) ensemble is insufficient in this regard
\cite{Mladek2007}.  Instead one has to estimate the lattice site or
cluster chemical potential $\mu_{\rm c}$, given by $N_{\rm c}\mu_{\rm c}=F+PV -\mu N$,
which vanishes at equilibrium \cite{Swope:1992fk}. Doing so entails
supplementing TI measurements of $F$ with additional sampling of the
chemical potential $\mu$ (via the Widom insertion method) and the
pressure $P$ (via the virial) \cite{Mladek2007}. This process, or
alternatively a direct estimation of the constrained free energy
\cite{Zhang:2010fk}, then has to be repeated for a range of values of
$n_{\rm c}$ in order to pinpoint equilibrium. Accordingly it is cumbersome and
laborious.

Here we introduce a new MC simulation scheme that allows direct
calculation of the chemical potential of crystals (and thence the free
energy) from a single simulation. Extending the method to cluster
crystals permits direct estimation of the cluster chemical potential, 
while histogram reweighting techniques can be used to identify the
equilibrium state without further simulation. We first describe the
basic scheme for a simple crystal before outlining its cluster solid--generalisation. 

The central idea is to construct (within the constant-$NPT$ ensemble) a
reversible sampling path between a lattice with $N+M$ particles and
another with $N$ particles. The relative probability of finding the
simulation in the two states provides a measure of the Gibbs free energy
difference $\Delta G=\mu M$. This yields the chemical potential $\mu$,
from which the Helmholtz free energy density follows immediately as
$f=\mu\rho-P$, with $P$ the prescribed pressure and $\rho$ the measured
particle number density. 

\begin{figure}[h]
\vspace*{-2mm}
\includegraphics[type=pdf,ext=.pdf,read=.pdf,width=0.85\columnwidth,clip=true]{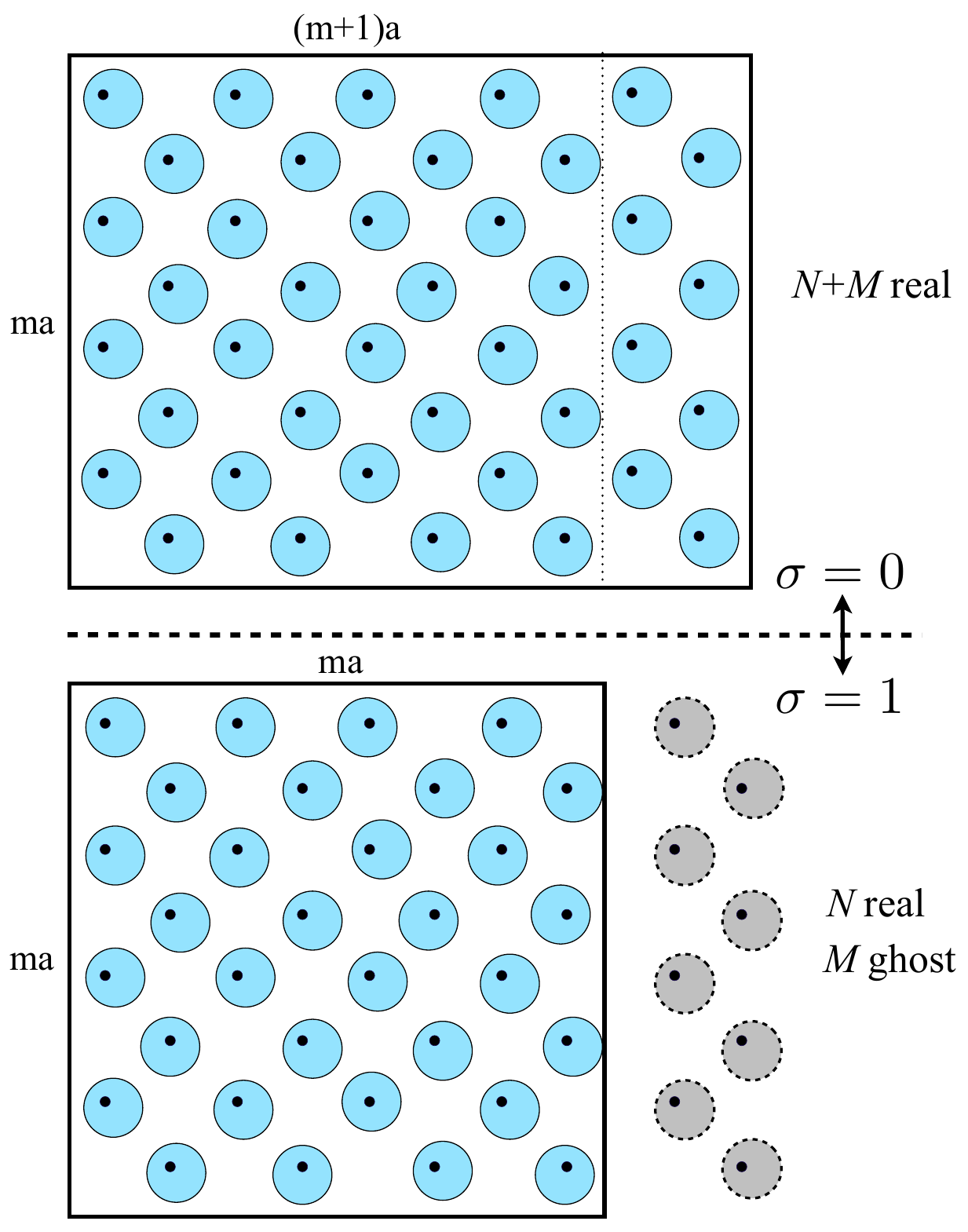}\\
\vspace*{-4mm}
\caption{(Color Online). Schematic of the ghost particle switching scheme described in the text.}
\label{fig:switching}
\end{figure}

To elaborate, consider the situation shown schematically in
Fig.~\ref{fig:switching} for a cubic lattice (though note the method is
applicable to any Bravais lattice). A constant-$NPT$ ensemble Monte Carlo
simulation \cite{Frenkelsmit2002} is to be found in one of two states
$\sigma\in\{0,1\}$. For $\sigma=0$ the system comprises a periodic box
of volume $V^{(0)}$ containing $(m+1)\times m\times m$ unit cells of lattice parameter $a$.
Each particle (a circle) is associated with a unique site of a fixed
perfect lattice (a black dot): there are $N$ particles in the cubic subvolume of
$m^3$ unit cells shown, and $M=N/m$ particles in the remaining (rightmost)
plane of unit cells. In the spirit of the phase switch method
\cite{Bruce1997a,Wilding2000}, we write the position vector of each
particle $i$ in terms of the displacement $\vec{u}_i$ from its lattice site
$\vec{R}^{(0)}_i$, i.e. $\vec{r}_i^{\:(0)}=\vec{R}_i^{(0)}+\vec{u}_i$, $i=1\ldots
N+M$.

The switch to the $\sigma=1$ state comprises a reversible mapping in
which the instantaneous particle displacements $\{\vec{u}_i\}$ are
re-associated with a second set of lattice sites $\{\vec{R}^{(1)}\}$, such that $
\vec{r}_i^{\:(1)}=\vec{R}_i^{(1)}+\vec{u}_i$. We set $\vec{R}_i^{(0)}=\vec{R}_i^{(1)}$ for
$i=1\ldots N$. Thus for $\sigma=1$, the first $N$ lattice sites form a
periodic cubic system with volume $V^{(1)}=V^{(0)}m/(m+1)$, and
the particles associated with these sites retain their relative positions within the box
under the switch. By contrast, for the
remaining $M$ particles, the change in environment under the switch is
more radical; they leave the box altogether to become ``ghost''
particles, associated with fixed sites $\{\vec{R}_i^{(1)}\},\:\; i=N+1\ldots
N+M$. Ghost particles are {\em independent} (so the relative positions
of the fixed sites is arbitrary) and they experience only a harmonic
confining potential $\phi_g(\vec{u}_i)$  whose amplitude is chosen to roughly
match the average ghost particle displacement to that of real particles.

Consider now the associated statistical mechanics. If we write the
partition function of the system in state $\sigma$ as $Z^{(\sigma)}$,
then since ghost particles are independent,
\bea
Z^{(0)}&=& Z(N+M,P,T)\:,\\
Z^{(1)}&=& Z_g^M Z(N,P,T)\:,
\eea
where $Z_g=\int d\vec{u}\exp(-\beta\phi_g(\vec{u}))$ is the (exactly
calculable) partition function of one ghost particle, with
$\beta=1/(k_{\rm B}T)$ as usual. The free energy change associated with the switch $\sigma=1\to 0$
follows as
\be
\beta \Delta G=\beta M\mu=\ln \frac{Z(N,P,T)}{Z(N+M,P,T)}=\ln \frac{Z^{(1)}}{Z^{(0)}Z_g^M},
\label{eq:mu}
\ee
with $\mu$ the chemical potential. In order to estimate $\Delta G$, we
supplement standard MC updates of the particle displacements and box
volume with attempts to switch  $\sigma$. These are accepted with
the standard ($NPT$) ensemble probability:  $p_a={\rm
min}[1,\alpha\exp{(-\beta(P\Delta V+\Delta E))}]$, with
$\alpha=V^{(1-\sigma)}/V^{(\sigma)}$. In general, however, such switch
attempts suffer low acceptance rates since the particle displacements
$\{\vec{u}_i\}$ for the current $\sigma$ may not all be typical for the
switched $\sigma$. To deal with this we implement biased (``umbrella'')
sampling \cite{Frenkelsmit2002}, which enhances the probability of
configurations (for each $\sigma$) for which the instantaneous switch
cost ${\kappa}=\beta(p\Delta V+\Delta E)$ is small. Specifically we
include a weight function $\eta_\sigma(\kappa)$ in the acceptance
probabilities for all MC updates. Weights are obtainable via any of the
standard techniques such as transition matrix or Wang Landau sampling.
Their effects are unfolded from the sampling in the usual way
\cite{Frenkelsmit2002} at the end of the simulation.

In this manner one accumulates the relative probability 
$p^{(1)}/p^{(0)}=Z^{(1)}/Z^{(0)}$ of finding the simulation in the
respective $\sigma$ states, from which the requisite chemical potential
follows as
\be
\mu=k_{\rm B}T[M^{-1}\ln (p^{(1)}/p^{(0)}) - \ln Z_g ]\:.
\ee
Statistical errors in $\mu$ are determined simply by the switching
statistics and readily quantified via a block analysis. If the
uncertainty in  $r=p^{(1)}/p^{(0)}$ is $\alpha$, then that in $\mu$ is
$O(\alpha/M)$. Since $M$ is typically $O(10^2)$, this bestows the
method with high sensitivity. To test it we have measured $\mu$ for
a Lennard-Jones fcc crystal of system size
$m=4$. Interactions were truncated at $r_c=2.9$ and a mean-field
correction of the usual type \cite{Frenkelsmit2002} was applied. As no
previous estimates of chemical potentials for this system exist (to our
knowledge), we use our results to calculate the absolute Helmholtz free
energy density $f=\mu\rho-P$, and compare with literature estimates
determined by TI. Only one such previous estimate quotes
an associated uncertainty,
allowing for meaningful comparison \cite{Vega:2007uq}. For the state
point $T=2.0,\rho=1.28$, we find $\beta P=20.985(2),\beta\mu=18.967(3),\beta
f=3.292(1)$, the latter comparing well with the estimate of Vega {\em et al.}
$\beta f=3.290(4)$, though our error bar is substantially smaller.

%\begin{table}[h]
%\begin{tabular}{|c c | c c c  c |}    \hline
% $T$    &  $\rho$  &  $\beta P$           &  $\beta\mu $ & $  \beta f  $ & $\beta f$ (TI)  \\ \hline
%$0.75$ &  $1.0$   &  $2.60(1)$     &  $ -3.15(2)  $ & $ -5.40(2) $ & $ -5.48 $  ref.\cite{kang:1986fk} \\
%$1.35$ &  $1.1$   &  $9.06(1)$     &  $ 7.51(2)   $ & $ -0.80(2) $ & $ -0.72 $  ref.\cite{kang:1986fk} \\
%$2.0$  &  $1.28$  &  $20.986(5)$   &  $ 18.965(10)$ & $  3.2866(10)$ & $ 3.290(4) $ ref.\cite{Vega:2007uq}    \\ 
%$4.0$  &  $1.389$ &  $22.1(1)$     &  $ 20.49(3)  $ & $  6.37(3) $ & $  6.407$  ref.\cite{Barroso:7145}    \\ 
%$5.0$  &  $1.4$   &  $20.3(12)$    &  $ 19.16(3)  $ & $  6.53(2) $ & $  6.57 $  ref.\cite{kang:1986fk}  \\ \hline
%\end{tabular} 
%\caption{Measured pressure, chemical potential and free energy density of
%the fcc LJ solid at a selection of state points.}
%\label{tab:LJ}
%\end{table}

We next address a challenging problem in the simulation of cluster
solids. At some $(T,\rho)$, equilibrium corresponds to a particular
value of the cluster occupancy $n_{\rm c}$ and the lattice parameter $a$. But
specifying $\rho$ fixes neither of these parameters, e.g.\ in an fcc
cluster crystal, $\rho=4n_{\rm c}/a^3$, which can be realized by many
combinations of $n_{\rm c}$ and $a$. In a real system, $N_{\rm c}$ changes to relax
the system to equilibrium. However, in conventional simulation
ensembles, the value of $N_{\rm c}$ is constrained on accessible times scales
by free energy barriers and does not fluctuate. As described above, one
approach to determining equilibrium in these circumstances is to
estimate the cluster chemical potential $\mu_{\rm c}(n_{\rm c})$ via a laborious
combination of TI, Widom particle insertion and virial sampling
\cite{Mladek2007}, while another is to directly estimate the constrained free
energy via TI \cite{Zhang:2010fk}. By contrast ghost particle switching (or more precisely ghost {\em
cluster} switching) provides a simpler, more efficient and elegant
solution to this problem.  

In seeking to apply the method, it is expedient 
\begin{comment} both
in terms of capitalizing on histogram reweighting and for directly
estimating the cluster chemical potential,
\end{comment}  
to employ an ensemble in which both $n_{\rm c}$ and $a$ are free to
fluctuate--the constant $\mu,P,T$ ensemble. This ensemble is rarely
utilized in simulations because the extensive scaling of the entropy
means that the partition function is finite for high pressures,
diverges on approach to equilibrium, e.g.\ $Z(\mu,P,T)\sim
(P-P_{\rm eq})^{-1}$ as $P\to P_{\rm eq}^+$, and is infinite for all lower pressures.
This is no longer a problem when we have a constraint of fixed
$N_{\rm c}$. The partition function is then
\[
Z(\mu,P,T,N_{\rm c}) = \sum_N \int dV\,dE\,e^{S(N,V,E,N_{\rm c})-\beta(E + PV -\mu N)}
\]
The entropy will be extensive for large $N_{\rm c}$, $S(N,V,E,N_{\rm c}) = N_{\rm c}
s(n_{\rm c},v_{\rm c},e_{\rm c})$ with $n_{\rm c}=N/N_{\rm c}$, $v_{\rm c}=V/N_{\rm c}$, $e_{\rm c}=E/N_{\rm c}$. The
dominant contribution to $Z$ comes from the maximum of the integrand,
where $-\beta\mu = \partial s/\partial n_{\rm c}$, $\beta P = \partial
s/\partial v_{\rm c}$ and $\beta= \partial s/\partial e_{\rm c}$. Denoting these
saddle point values with an asterisk, the extensive contribution to
$\ln Z$ is
\be
\ln Z(\mu,P,T,N_{\rm c}) = N_{\rm c} [s(n_{\rm c}^*,e_{\rm c}^*,v_{\rm c}^*)
-\beta(e_{\rm c}^*-\mu n_{\rm c}^* + Pv_{\rm c}^*)]
\label{Z_Nc_saddle}
\ee
In general, this is a non-equilibrium partition function because at
equilibrium any two of $(\mu,P,T)$ determine the third. To find the
equilibrium condition, assume $(\mu,P,T)$ is a set of equilibrium
parameters, then so must $n_{\rm c}^*$, $v_{\rm c}^*$ and $e_{\rm c}^*$ be. This means that they
are obtained by maximizing the entropy $S(N,V,E,N_{\rm c})$ over $N_{\rm c}$. From
the extensive form of $S$ above, and using that
$\beta=\partial s/\partial e_{\rm c}$ etc, this shows directly that
the combination in square brackets in (\ref{Z_Nc_saddle}) must vanish.

The upshot of this analysis is that $-k_{\rm B}T\ln Z(\mu,P,T,N_{\rm c})/N_{\rm c}$
vanishes at equilibrium, which is as expected given that by standard
thermodynamic arguments this quantity can also
be identified with the cluster chemical potential $\mu_{\rm c}$.
On the other hand, $Z(\mu,P,T,N_{\rm c})$ is also the weight of different
$N_{\rm c}$ values in a simulation where $N_{\rm c}$ can fluctuate, and so the
above equilibrium criterion tells us that at equilibrium these weights
are to leading order independent of $N_{\rm c}$.
Ghost cluster switching allows one to repeatedly add and remove a
crystallographic plane of lattice sites, thereby circumventing the
barriers between different values of $N_{\rm c}$. Measurements of the relative
probabilities of macrostates with different $N_{\rm c}$ then directly
probes $\mu_{\rm c}$.

The implementation is technically similar to that described for simple
crystals, except that the particle number is permitted to fluctuate via
insertions/deletions which for cluster solids are efficient
owing to the lack of a repulsive hard core in the potential.  For ghost
sites, in addition to choosing the harmonic amplitude such that ghost
particle displacements are similar to those of real particles, we impose
a ghost chemical potential $\mu_g$ chosen to yield an average site
occupancy close to that of real sites. The order parameter against which
we bias to enhance the switch probability is extended from the single
occupancy case to become $\kappa=\beta(P\Delta V+\Delta E \pm
(\mu-\mu_g)N_g )$, with $N_g$ the instantaneous number of particles
associated with lattice sites $i=N+1\ldots N+M$. We sample the distribution of the observables
($N,V,E$) across the two values of $\sigma$, from which we unfold the
effects of the biasing weights and the ghost particle free energy. This
yields ({\it inter alia}), the volume distribution $p(V|\mu,P,T)$, which
exhibits two peaks, one corresponding to $\sigma=1$ and the other for
$\sigma=0$. Equilibrium is signaled in a very simple fashion by the
equality of the peak weights of $p(V)$, as these have the same ratio as
$Z(\mu,P,T,N_{\rm c}=N)$ and $Z(\mu,P,T,N_{\rm c}=N+M)$.

\begin{figure}[h]
\vspace*{-2mm}
\includegraphics[type=pdf,ext=.pdf,read=.pdf,width=0.8\columnwidth,clip=true]{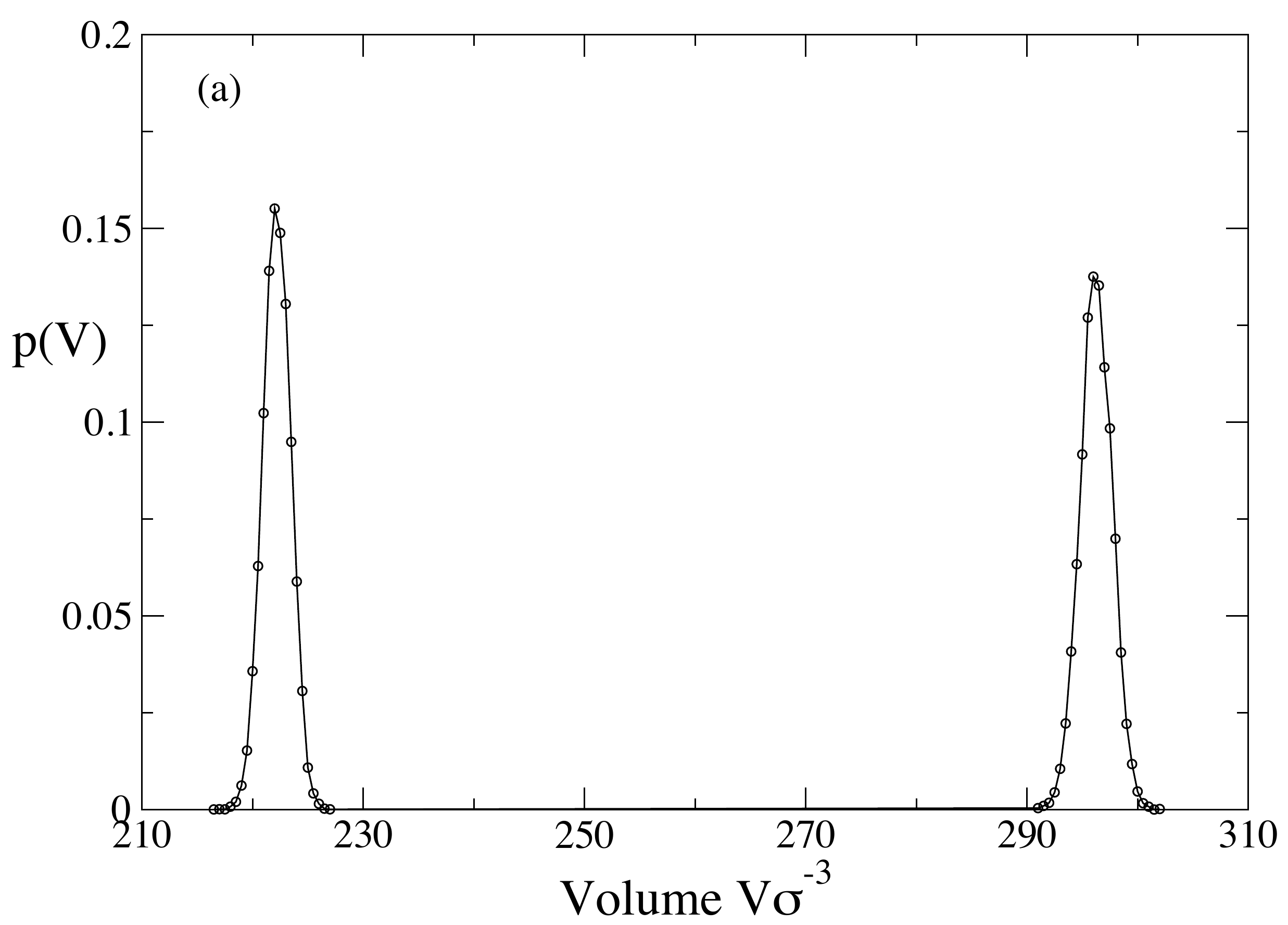}\\[2mm]
\includegraphics[type=pdf,ext=.pdf,read=.pdf,width=0.6\columnwidth,clip=true]{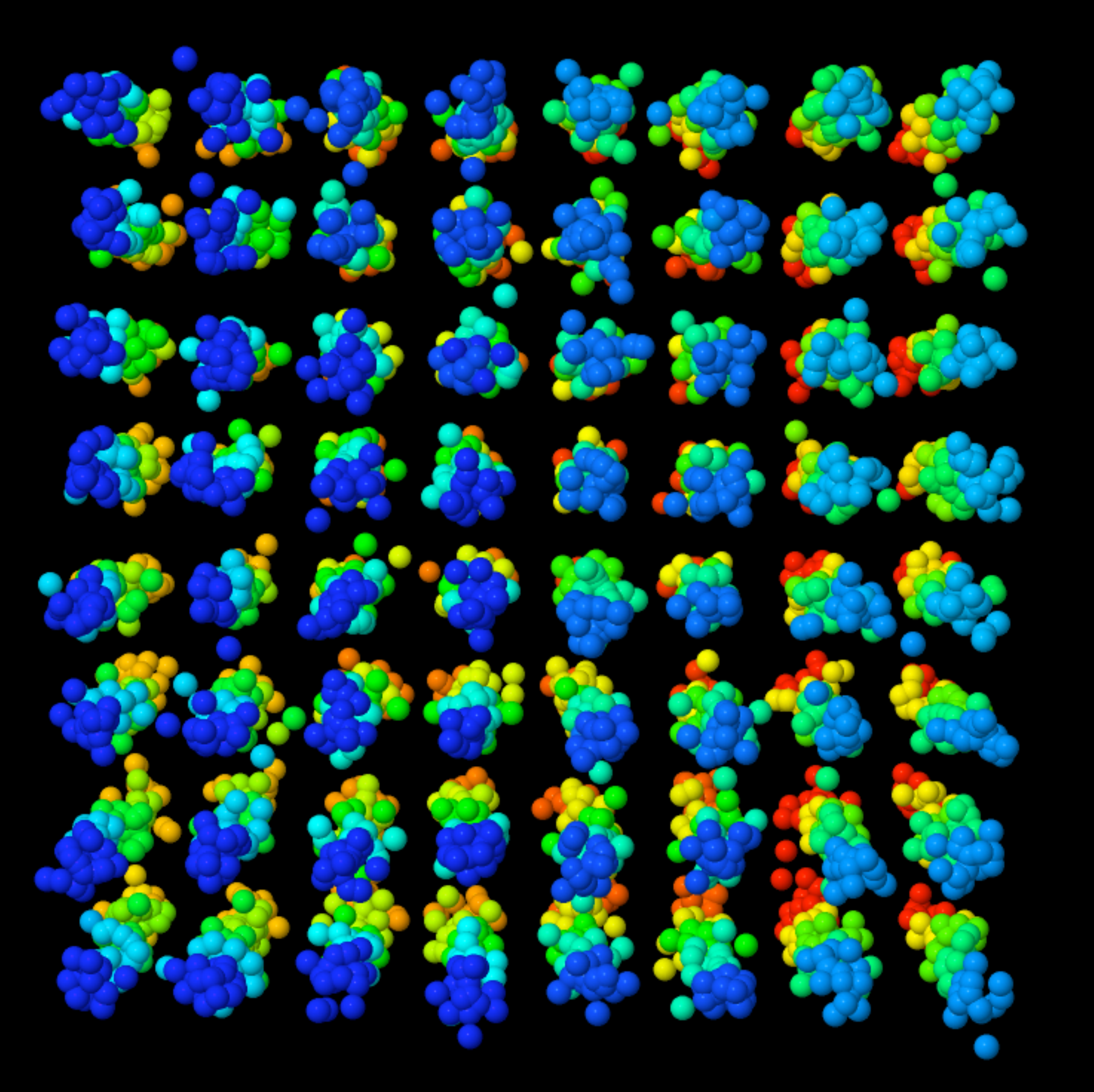}\\

\vspace*{-3mm} 

\caption{(Color Online). {\bf (a)} Equality of peak weights in $p(V)$ 
	that signifies equilibrium in the ($\mu,P,T$) ensemble (see text
for the equilibrium parameters). {\rm (b)} An equilibrium configuration
viewed along the [100] direction.}
\label{fig:findequil}
\end{figure}

To validate the methodology we have considered a prototype cluster
solid: the generalized exponential model (GEM-4) whose interaction potential
is $u(r)=\epsilon\exp((-r/\sigma)^4)$. We determined $P_{\rm eq}$ and
$\mu_{\rm eq}$ for $T=1.1, \rho=8.5$ in the fcc phase -- a state point for
which prior TI data is available \cite{Mladek2007}. The initial
simulation was performed for $P=114.5$, $\mu=29.7$ and the resulting
probability distribution $p(V)$ was subsequently extrapolated in $P$ and
$\mu$ to yield both equal peak weights and a density matching the target
value. Fig.~\ref{fig:findequil} shows the resulting equilibrium form,
obtained for $P_{\rm eq}=114.45(1)$, $\mu_{\rm eq}=29.752(3)$. The associated
cluster number is $n_{\rm c}=17.470(5)$ and the fcc lattice parameter
$a/\sigma=2.018(1)$. We note that these results agree to within error
with those of Mladek~\cite{Mladek2012}. Fig.~\ref{fig:findequil} shows a
portion of a snapshot of an equilibrium configuration colored by cluster. 

As an application we consider a recent proposal \cite{Neuhaus:2011fk}
concerning the existence of a cascade of low temperature critical points
in the GEM-4 model. On the basis of ground state energy calculations and
a phonon analysis, these authors found a sequence of low temperature
isostructural (fcc) phase transitions on increasing density. At $T=0$,
each transition is such that the cluster number changes by
unity from one integer value to the next, i.e.\ $n_{\rm c}\to n_{\rm c}+1$. No
theoretical evidence was found that any of these transition has an
associated critical point at finite temperature, but the authors
hypothesized that this should be the case.  Indeed  subsequent evidence
for a critical point terminating  the lowest density transition
$n_{\rm c}=2\leftrightarrow 3$ has been found using TI
\cite{Zhang:2010fk,Zhang:2012aa}. 

\begin{figure}[h]
\vspace*{-3mm}
\includegraphics[type=pdf,ext=.pdf,read=.pdf,width=0.8\columnwidth,clip=true]{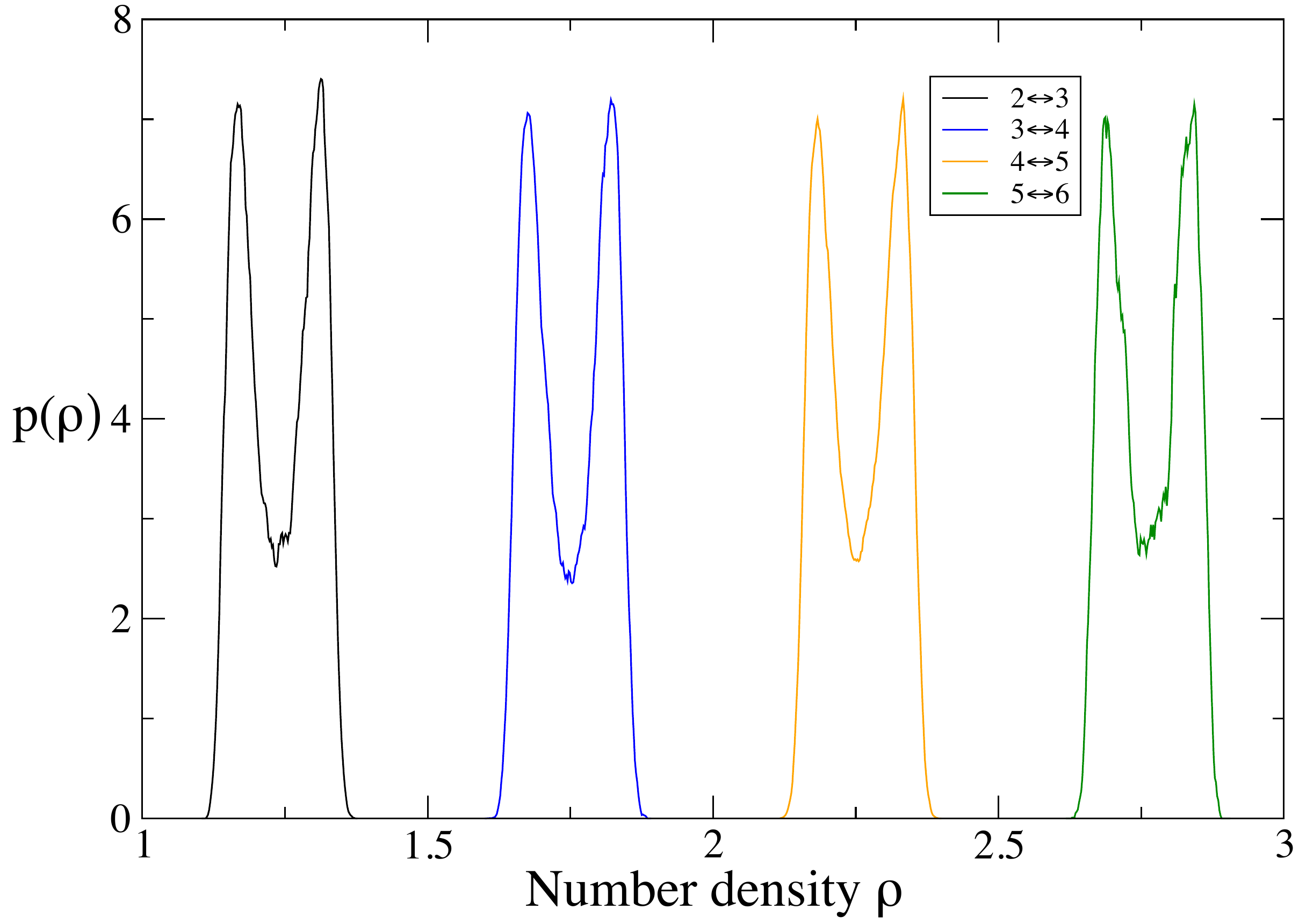}\\

\vspace*{-3mm}
\caption{(Color Online). Density distributions of the GEM-4 model corresponding to the
near-critical point parameters listed in table~\ref{tab:critpars}.}
\label{fig:densdists}
\end{figure}

We have used the present method to search for further critical points in
the model at higher density. We employed the known universal Ising form
of the critical order parameter distribution to estimate the first
four critical points \cite{Bruce1992}. The resulting critical point
parameters are listed in table~\ref{tab:critpars}. Surprisingly we find
that $T^{\rm c}$ is equal within error in each case as reflected in the
independence of the form of the density distributions at
$T=0.04348$ shown in Fig.~\ref{fig:densdists}.

\begin{table}[h]
	\begin{tabular}{|c|c c c c|}\hline
		{\rm Transition} & $\rho^{\rm c}$ & $P^{\rm c}$ & $\mu^{\rm c}$ & $T^{\rm c}$\\ \hline
		$2\leftrightarrow 3$ & $1.239(4)$ & $1.974(2)$ & $2.9151(5)$  & $0.0435(4)$\\ 
		$3\leftrightarrow 4$ & $1.740(4)$  & $3.879(2)$ & $4.1878(4)$ & $0.0435(4)$\\
		$4\leftrightarrow 5$ & $2.257(5)$ & $6.418(2)$ & $5.4521(5)$  & $0.0435(4)$ \\
 		$5\leftrightarrow 6$ & $2.762(5)$ & $9.575(3)$ & $6.709(5)$  & $0.0435(5)$\\\hline
		\hline
	\end{tabular}

\caption{Estimated values of the critical density $\rho^{\rm c}$, pressure
$P^{\rm c}$, chemical potential $\mu^{\rm c}$ and temperature $T^{\rm c}$, for
the first four members of the infinite cascade of phase transitions
having $n_{\rm c}\leftrightarrow n_{\rm c}+1$ at $T=0$.}

\label{tab:critpars}
\end{table}

In summary we have introduced an efficient and accurate `ghost particle
switching' method for chemical potential determination in crystalline
solids within the constant-$NPT$ ensemble. The method, which circumvents
the need for integration to distant references states and its attendant
pitfalls, requires only a single simulation at the state point of
interest and yields statistical uncertainties  directly and
transparently. Such access to the chemical potential permits the direct
determination of phase boundaries by matching of $\mu$ and $P$ in the
coexisting phases.  An extension of the method to multiple occupancy
crystals simplifies the problem of determining their
equilibrium parameters.  As a demonstration of its power in this regard,
we have studied the GEM-4 cluster solid, uncovering the presence of
a cascade of critical points. More generally, the basic approach of
ghost particle switching should be applicable to any system exhibiting
periodic microphase separation such as lamellar or micellar crystals
\cite{Alexandridis:2000}, where the repeat unit can
contain many individual particles. 

%\acknowledgments 
%Computational results were produced on a machine funded by HEFCE’s
%Strategic Research Infrastructure fund.

%TC:break biblio
\bibliography{Papers}
\bibliographystyle{prsty}
%TC:break supplementary

\end{document}